\documentclass[twocolumn,aps,showpacs,nofootinbib,prd]{revtex4-1}
\usepackage{amssymb}
\usepackage{amsmath}
\usepackage{graphics}
\usepackage{epsfig}
\usepackage[dvips]{color}
\usepackage{float}
\usepackage{soul}
\newcommand{\be}{\begin{equation}}
\newcommand{\ee}{\end{equation}}
\newcommand{\bea}{\begin{eqnarray}}
\newcommand{\eea}{\end{eqnarray}}
\newcommand{\beas}{\begin{eqnarray*}}
\newcommand{\eeas}{\end{eqnarray*}}

\newcommand{\nn}{\nonumber\\}
\newcommand{\slsh}[1]{{\not \! #1}}

\begin{document}
\title{Thermomagnetic correlation lengths of strongly interacting matter in the Nambu--Jona-Lasinio model}
\author{Alejandro Ayala$^{1,2}$, L. A. Hern\'andez$^1$, M. Loewe$^{2,3,4}$, Alfredo Raya$^5$ J. C. Rojas$^6$, R. Zamora$^{7,8}$}
  \address{
  $^1$Instituto de Ciencias
  Nucleares, Universidad Nacional Aut\'onoma de M\'exico, Apartado
  Postal 70-543, M\'exico Distrito Federal 04510,
  Mexico.\\
  $^2$Centre for Theoretical and Mathematical Physics, and Department of Physics,
  University of Cape Town, Rondebosch 7700, South Africa.\\
  $^3$Instituto de F\1sica, Pontificia Universidad Cat\'olica de Chile,
  Casilla 306, Santiago 22, Chile.\\
  $^4$Centro Cient\1fico-Tecnol\'ogico de Valpara\1so, Casilla 110-V, Valpara\1so, Chile.\\
  $^5$Instituto de F\1sica y Matem\'aticas, Universidad Michoacana de San Nicol\'as de Hidalgo, 
  Edificio C-3, Ciudad Universitaria, Morelia, Michoac\'an 58040, Mexico.\\
  $^6$Departamento de F\1sica, Universidad Cat\'olica del Norte, Casilla 1280, Antofagasta, Chile,\\
  $^7$Instituto de Ciencias Basicas, Universidad Diego Portales, Casilla 298-V, Santiago, Chile,\\
  $^8$Centro de Investigaci\'on y Desarrollo en Ciencias Aeroespaciales (CIDCA), Fuerza A\'erea de Chile, Santiago, Chile.}

\begin{abstract}

We study the correlation length between test quarks with the same electric and color charges in the Nambu--Jona-Lasinio model, considering thermal and magnetic effects. We extract the correlation length from the quark correlation function. The latter is constructed from the probability amplitude to bring a given quark into the plasma, once a previous one with the same quantum numbers is placed at a given distance apart. For temperatures below the transition temperature, the correlation length starts growing as the field strength increases to then decrease for large magnetic fields. For temperatures above the pseudocritical temperature, the correlation length continues increasing as the field strength increases. We found that such behavior can be understood as a competition between the tightening induced by the classical magnetic force versus the random thermal motion. For large enough temperatures, the increase of the occupation number contributes to the screening of the interaction between the test particles. The growth of the correlation distance with the magnetic field can be understood as due to the closer proximity between one of the test quarks and the ones popped up from vacuum, which in turn appear due to the increase of the occupation number with temperature. 

\end{abstract}

\maketitle

\section{Introduction}	\label{secI}

The response of strongly interacting matter to the effects of magnetic fields at finite temperature has received a great deal of attention in recent years. Physical scenarios where the presence of these fields may be relevant include semicentral collisions of heavy nuclei at high energy and the interior of compact stellar objects. Research in this area has been stimulated by the lattice QCD (LQCD) result, dubbed {\it inverse magnetic catalysis} (IMC), whereby the pseudocritical temperature for the chiral/deconfinement phase transition and the quark-antiquark condensate for temperatures above the pseudocritical temperature decrease with increasing field strength~\cite{bali1,bali2,bali3}. 

In order to grasp the microscopic origin of IMC it is possible to resort to effective QCD models. It turns out that, within a mean field approach, these models do not explain IMC~\cite{Boomsma01,Loewe1,Fraga,Fraga2,Blaschke01}; extra ingredients seem to be called for. Recall that a working explanation for IMC, borne out of an analysis of the full QCD partition function, is given in terms of the competition between the so called valence and sea quark contributions to the condensate~\cite{bali1,bali2,Bruckmann}: In a magnetized medium, the sea contribution becomes important around the transition temperature and is opposite to that of the valence contribution. Since the former can be viewed as a back reaction of fermions on the gauge fields and these last are absent in effective models, the common lore is that these models require additional effects in order to capture such competition. 

Another QCD based approach to IMC is discussed in Refs.~\cite{Ayala1}, where the one-loop effective quark-gluon coupling at zero and  finite temperature is computed in the presence of a magnetic field. It is shown  that at zero (high) temperature the coupling grows (decreases) with increasing magnetic field strength. The effect is due to a subtle competition between the color charges of gluons and quarks. At zero temperature, the former is larger than the latter whereas at high temperature, the coupling receives contributions only from the color charge associated to quarks. It seems therefore that a competition between opposite effects whose strength varies as the temperature increases is essential to describe IMC.

It has also been found that when effective models include temperature and magnetic field-dependent couplings, IMC can be explained~\cite{decreasing1}. Most notably, it has been realized that these modified couplings can be computed self-consistently in calculations beyond the mean field approximation~\cite{decreasing2}. The picture that emerges is that effective models are able to capture some of the dynamical aspects of the full theory by allowing the running of couplings with both the temperature and the magnetic field strength. 

A step forward in the search for the thermomagnetic properties of effective model parameters has been given in Refs.~\cite{NJL1, NJL2}. Using a reverse engineering approach, whereby LQCD data for the quark condensate in the presence of a magnetic field at finite temperature is described using the Nambu--Jona-Lasinio model, it has been shown that it is possible to extract a non-trivial behavior of the coupling $G$ and dynamical generated mass $M$ as functions of the magnetic field strength and temperature. Such behavior provides a better description of several thermodynamical quantities than the calculation without a thermomagnetic dependence of the parameters, in particular of the LQCD longitudinal and transverse components of the pressure~\cite{NJL2}. 

It has been suggested that the properties of the coupling are reminiscent of asymptotic freedom, namely, that the interaction strength decreases as the energy scale increases. In this picture, when the system is above the chiral/deconfinement phase transition temperature, an increase of the field strength speeds up the running of the coupling towards weaker values. Asymptotic freedom can also be viewed in terms of the spatial properties of the strongly interacting system. Therefore, it is desirable to round up the picture and see how the running of the parameters with temperature and field strength translates into the change of the correlation distance between test color and electric charges. In this work we set up to answer this question. We use the results obtained in Ref.~\cite{NJL2} to compute the correlation function between quarks with the same quantum numbers and from this we extract how their distance, both in the transverse and longitudinal directions with respect to the magnetic field, change as a function of the temperature and field strength.

The work is organized as follows: In Sec.~\ref{II}, we briefly recall the results of Ref.~\cite{NJL2} for the thermomagnetic behavior of the coupling $G$ and dynamically generated mass $M$. We show that this behavior does reproduce the magnetic field dependence of the pseudocritical temperature. In Sec.~\ref{III} we compute the correlation function for quarks with the same quantum numbers. In Sec.~\ref{IV} we extract the correlation lengths and study their thermomagnetic properties. Finally, in Sec.~\ref{concl} we summarize and conclude.

\section{Pseudocritical temperature}\label{II}

Effective theories are one possible approach to study strongly interacting matter under the influence of a magnetic field. One of such theories is the Nambu-Jona-Lasinio model (NJL), whose Lagrangian density in the mean field approximation is written as
\begin{equation}
\mathcal{L}_{MF}=-\frac{\sigma^2}{4G}+\bar{\psi}(i\slsh{\partial}-M)\psi,
\end{equation}
where $\sigma = 4G\langle\bar\psi\psi\rangle$ and $M=m+\sigma$. In Ref.~\cite{NJL2}, some of us used this model to study the thermomagnetic properties of the dynamically generated mass $M$ and the coupling $G$. For that purpose, we resorted to relate the LQCD results for the light-quark condensate~\cite{bali2}, $\langle \bar{\psi} \psi \rangle$, as a function of $T$ and $B$ to $M$ and $G$ by means of the \textit{gap equation}
\begin{equation}
 M-m=4G\int\frac{d^4p}{(2\pi)^4}\text{Tr}[iS(p)],
 \label{gap1}
\end{equation}
where the light-quark condensate is given by
\begin{equation} 
 \langle \bar{\psi}\psi\rangle = -\int\frac{d^4p}{(2\pi)^4}\text{Tr}[iS(p)].
 \label{qbarq}
\end{equation}
The effect of the magnetic field in Eq.~(\ref{qbarq}) is reflected in the dressing of the quark propagator. For this we use Schwinger's proper time representation of the  two-point function
 \begin{eqnarray}
  && i S(p) = \int _{0}^{\infty}\frac{ds}{\cos(q_fBs)}e^{is(p_{\parallel}^{2} - p_{\perp}^{2}\frac{\tan (q_fBs)}   {q_fBs}- M^{2} +i\epsilon)}\nonumber\\
  &&\biggl[\left(\cos (q_fBs) + \gamma_1 \gamma_2 \sin (q_fBs)\right)
   (M+\slsh{p_{\|}}) - \frac{\slsh{p_\bot}}{\cos(q_fBs)} \biggr],\nonumber\\
   \label{tracewithSchwinger}
 \end{eqnarray}
where $q_f$ is the absolute value of the quark charge (i.e. $q_u = 2|e|/3$ and $q_d = |e|/3$), and we have chosen the homogeneous magnetic field to point in the 
$\hat{z}$ direction, namely $\boldsymbol{B}=B\hat{z}$. This configuration can be obtained from an external vector potential which we choose in the so called {\it symmetric gauge}
\begin{equation}
A^{\mu}= \frac{B}{2}(0,-y,x,0).
\end{equation}
We have also defined
\bea
p_\perp^\mu&\equiv&(0,p_1,p_2,0),\nonumber\\
p_\parallel^\mu&\equiv&(p_0,0,0,p_3),\nonumber\\
p_\perp^2&\equiv& p_1^2+p_2^2,\nonumber\\
p_\parallel^2 &\equiv& p_0^2-p_3^2.
\label{pps}
\eea
Notice that since the magnetic field breaks Lorentz invariance, the propagator involves a non-local, albeit path independent phase. However, by taking the trace over a closed one-loop diagram, as is required for the calculation of the condensate, this phase does not contribute and thus we ignore it in the sequel.

Also, in order to introduce a finite temperature within the Matsubara formalism, we transform the integrals to Euclidean space by means of
\begin{equation}
\int \frac{d^{4}p}{(2\pi)^{4}} f(p) \to iT \sum_{n= -\infty}^{+\infty}\int \frac{ d^3p}{(2\pi)^3} f(i\tilde{\omega}_n, \textbf{p}),
\label{tang}
\end{equation}
where the integral over the zeroth component of the fermion momentum has been discretized and we introduced the fermion Matsubara frequencies $\tilde{\omega}_{n} = (2n+1)\pi T$.

\begin{table}[t!]
\centering
\begin{tabular}{cccccc}
\hline
  $\tau_0$ & $-\langle\bar{\psi}\psi\rangle_0^{1/3}$ & $M_0$ & $G_0$ & $m$ & $T_c^{NJL}$\\
  (GeV)$^{-2}$ & (GeV) &  (GeV) & (GeV)$^{-2}$ &  (GeV) & (GeV) \\ \hline
 1.27 & 0.220 & 0.224 & 5.08 & 0.00758 & 0.267 \\ \hline
 0.74 & 0.260 & 0.192 & 2.66 & 0.00465  & 0.228
\end{tabular}
\caption{Two sets of values for the vacuum regulator $\tau_0$, condensate $\langle\bar{\psi}\psi\rangle_0$ and dynamically generated mass $M_0$ stemming from requiring that the pion mass and the pion decay constant computed in the NJL model attain their physical values. Shown also are the corresponding vacuum values for the coupling constant $G_0$, current quark mass $m$ and the pseudocritical temperature for $eB=0$.}
\label{table1}
\end{table}

Since the NJL model is non-renormalizable, the integral in Eq.~(\ref{qbarq}) needs to be regularized. The procedure involves isolating the vacuum piece, which is the only divergent term. The thermomagnetic contribution turns out to be finite and the result is
\begin{eqnarray}
   \langle \bar{\psi}\psi\rangle_{B,T} &=& -\frac{N_{c} M}{4\pi ^{2}}\frac{1}{2}\sum_{f} 
   \Big\{
   \int _{0}^{\infty}\frac{d\tau}{\tau^2}e^{-\tau M^{2}}\nonumber\\
   &\times&\left[\frac{q_fB\tau}
   {\tanh(q_fB\tau)}-1\right]
   \nonumber\\
&+&
   2 q_fB  \sum_{n=1}^{\infty}(-1)^n 
   \int _{0}^{\infty}d\tau \frac{e^{-\tau M^{2}}e^{-\frac{n^{2}}{4\tau T^{2}}}}{\tau \tanh(q_fB\tau)}\Big\}, 
\label{condpropthermmag}
\end{eqnarray}
whereas the vacuum contribution is given by
\bea
   \langle \bar{\psi}\psi\rangle_0=-\frac{N_cM_0}{4\pi^2}
   \int_{\tau_0}^\infty\frac{d\tau}{\tau^2}e^{-\tau M_0^2}.
\label{vaccond}
\eea
The quantity $M\equiv M(B,T)$ in Eq.~(\ref{condpropthermmag}) is such that when $B,T\to 0$, $M\to M_0$. It turns out that the integrals in Eq.~(\ref{condpropthermmag}) are finite as the lower limit of integration goes to zero. This means that the thermo-magnetic effects are independent of the regulator $\tau_0$ and we have consequently set it to zero in Eq.~(\ref{condpropthermmag}).

To fix the vacuum values of the light-quark condensate and the dynamical generated mass $M_0$, we choose the ultraviolet cutoff $\tau_0$ such that the model reproduces the physical values of the pion mass and of the pion decay constant.  Table~\ref{table1} shows two possible sets of parameters~\cite{Norberto}.

To explore some of the consequences of the extracted behavior of $M$ and $G$ as functions of the temperature and field strength, in Ref.~\cite{NJL2} we computed the thermomagnetic contribution to the longitudinal and transverse pressures. We found that below $T_c$, the transverse pressure as a function of the magnetic field decreases towards negative values, starting off from zero. However, for temperatures above the transition temperature, although the transverse pressure still decreases as a function of the field strength, it starts off from positive values. This turnover behavior of the transverse pressure means that above $T_c$ particles are pulled closer together, at least for small values of the magnetic field. The fact that at the same time the coupling decreases can be viewed as signaling that the strength of the bound of the condensate is smaller and this behavior can be responsible for the decrease of the condensate as the magnetic field strength is turned on.

\begin{figure}[t!]
 \centering
  \includegraphics[width=0.48\textwidth]{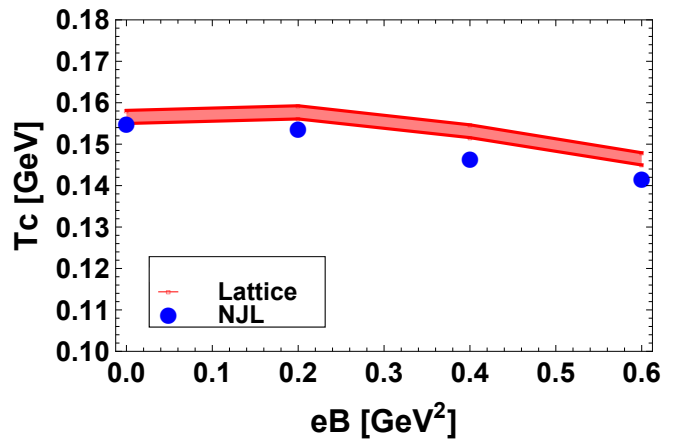}
  \caption{Pseudoritical temperature as a function of the magnetic field strength. The band corresponds to the LQCD values of Ref.~\cite{bali2} and the dots denote the NJL results. $M_0=224$ MeV}
\label{fig1}
\end{figure}

Another important quantity that can be studied from our results is the behavior of the pseudocritical temperature as a function of the field strength. The pseudocritical temperature can be obtained as the value of $T$ that maximizes the, {\it chiral succeptibility} $\chi_{B,T}$, namely, the (minus) derivative of the quark condensate with respect to the temperature, {\it i.e.}, the value of $T$ that satisfies
\begin{equation}
\chi_{B,T}\equiv
-\frac{\partial}{\partial T } \langle \bar{\psi} \psi  \rangle_{B,T} =0.
\label{critica}
\end{equation}
The solution of Eq.~(\ref{critica}) is found by means of Eq.~(\ref{qbarq}) together with the extracted values for $M(T,B)$. The result is shown in Fig.~\ref{fig1} where we also plot the corresponding values obtained by the LQCD calculation of Ref.~\cite{bali2}.  Notice that the result of the calculation for the pseudocritical temperature accounts for the fact that $M$ is a function of $T$ and $B$. 

It is important to notice that the pseudocritical temperature given by the model depends on the value of $M_0$ and does not coincide with the corresponding value reported by LQCD. It is therefore necessary to scale the temperatures obtained in the model to make them correspond to the physical values. Following our earlier work~\cite{NJL2}, the simplest choice is a linear scaling such that
\bea
   T^{NJL}=\left( \frac{T_c^{NJL}}{T^{LQCD}_c} \right) T,
\label{scale}
\eea
where $T^{LQCD}_c = 158$ MeV corresponds to the LQCD pseudocritical temperature. For all of the calculations we take $M_0=224$ MeV. There is very little dependence of the results when one takes the second choice for $M_0$ in Table~\ref{table1}.

In order to round up the picture and to have a better qualitative grasp of the origin of IMC, we now proceed to study the correlation function between two test quarks with the same quantum numbers in coordinate space.  

\section{Correlation function}\label{III}

In order to explore how the correlation distance between test (color and electric) charges is modified when considering thermomagnetic effects, we look at the correlation function. This object is constructed from the probability amplitude to place a test quark with quantum numbers $b$ (spin projection), $q$ (electric charge) and $\alpha$ (color charge) in position $\vec{x}'$ in a state that already contains a particle with the same quantum numbers but placed in position $\vec{x}$. In order to keep track of the indices, let us focus on the spin projection indices and take the other quantum numbers already to be the same. Furthermore, we first consider that the original particle is labeled with the spin projection $a$. 
Such amplitude is given by
\bea
   A(\vec{x},\vec{x}')=\langle\psi_a(\vec{x})\psi_b(\vec{x}')\bar{\psi}_b(\vec{x}')\bar{\psi}_a(\vec{x})\rangle,
\label{density}
\eea
where a sum over $a$ and $b$ is implicit. We use that
\bea
   \bar{\psi}_b(\vec{x}')&=&\psi^\dagger_c(\vec{x}')\gamma^0_{cb}\nn
   \bar{\psi}_a(\vec{x})&=&\psi^\dagger_d(\vec{x})\gamma^0_{da},
\label{barpsi}
\eea
to write
\bea
    A(\vec{x},\vec{x}')=
    \langle\psi_a(\vec{x})\psi_b(\vec{x}')\psi^\dagger_c(\vec{x}')\psi^\dagger_d(\vec{x})\rangle\gamma^0_{cb}\gamma^0_{da}.
\label{rewrite}
\eea
Recall that the equal-time commutation relations for the fermion fields are
\bea
   \{\psi_a(\vec{x}),\psi_b^\dagger(\vec{x}')\}&=&\delta_{ab}\delta^3(\vec{x}-\vec{x}')\nn
   \{\psi_a(\vec{x}),\psi_b(\vec{x}')\}&=&\{\psi_a^\dagger(\vec{x}),\psi_b^\dagger(\vec{x}')\}=0.
\label{commutation}
\eea
Using these commutation relations we have, on the one hand, when commuting $\psi^\dagger_d(\vec{x})$ to place it to the right of $\psi_a(\vec{x})$,
\bea
   A(\vec{x},\vec{x}') &=&
   \left[\langle\psi_a(\vec{x})\psi_d^\dagger(\vec{x})\psi_b(\vec{x}')\psi_c^\dagger(\vec{x}')\rangle\right.\nn
   &-&
   \left.\frac{}{}\delta_{bd}\delta^3(\vec{x}-\vec{x}')\langle\psi_a(\vec{x})\psi^\dagger_c(\vec{x}')\rangle\right]
   \gamma^0_{cb}
   \gamma^0_{da}.
\label{using1}
\eea
On the other hand, when commuting $\psi^\dagger_c(\vec{x}')$ to place it to the right of $\psi_a(\vec{x})$, we get
\bea
   A(\vec{x},\vec{x}') &=&-
   \left[\langle\psi_a(\vec{x})\psi_c^\dagger(\vec{x}')\psi_b(\vec{x}')\psi_d^\dagger(\vec{x})\rangle\right.\nn
   &+&
   \left.\frac{}{}\delta_{bc}\delta^3(\vec{x}-\vec{x}')\langle\psi_a(\vec{x})\psi^\dagger_d(\vec{x})\rangle\right]
   \gamma^0_{cb}
   \gamma^0_{da}.
\label{using2}
\eea
The second term on the right-hand side of Eq.~(\ref{using1}) vanishes identically except when $\vec{x}=\vec{x}'$, where it diverges. Since we are looking for the correlation when varying the distance between the test particles, we discard such contribution. Also, the second term  on the right-hand side of Eq.~(\ref{using2}) is proportional to the trace of $\gamma^0$, which vanishes. Therefore, adding up Eqs.~(\ref{using1}) and~(\ref{using2}), we get
\bea
   A(\vec{x},\vec{x}') &=&\frac{1}{2}\left[
   \langle\psi_a(\vec{x})\bar{\psi}_a(\vec{x})\psi_b(\vec{x}')\bar{\psi}_b(\vec{x}')\rangle\right.\nn
   &-&\left. \langle\psi_a(\vec{x})\bar{\psi}_b(\vec{x}')\psi_b(\vec{x}')\bar{\psi}_a(\vec{x})\rangle
   \right].
\label{sumusing12}
\eea
We now assume that the correlations in both terms of the right-hand side of Eq.~(\ref{sumusing12}) {\it factorize} to write
\bea
   A(\vec{x},\vec{x}') &=&\frac{1}{2}\left[
   \langle\psi_a(\vec{x})\bar{\psi}_a(\vec{x})\rangle\langle\psi_b(\vec{x}')\bar{\psi}_b(\vec{x}')\rangle\right.\nn
   &-&\left. \langle\psi_a(\vec{x})\bar{\psi}_b(\vec{x}')\rangle\langle\psi_b(\vec{x}')\bar{\psi}_a(\vec{x})\rangle
   \right].
\label{factorize}
\eea
The factorization assumption is a reasonable approximation for statistical systems in thermal equilibrium. Taking the spin projection indices $a$ and $b$ to be equal, we get
\bea
   A(\vec{x},\vec{x}') &=&\frac{1}{2}\left[
   \left({\mbox{Tr}}\left[S(0)\right]\right)^2-\left({\mbox{Tr}}\left[S(\vec{x}'-\vec{x})\right]\right)^2
   \right],
\label{equalindices}
\eea
where
\bea
   S(\vec{x}'-\vec{x})=\langle\bar{\psi}(\vec{x}')\psi(\vec{x})\rangle,
\label{spaceprop}
\eea
is the quark propagator in coordinate space at equal times. Notice that, since the order of the coordinates in the two factors of the second term on the right-hand side of Eq.~(\ref{factorize}) is reversed, the phase factor cancels. Since the trace of the Fourier transform of the translationaly invariant part of the propagator turns out to be real, hereafter $S(\vec{x}'-\vec{x})$ is just meant to represent the Fourier Transform of Eq.~(\ref{tracewithSchwinger}) at equal times. 

Dividing by the first term on the right-hand side of Eq.~(\ref{spaceprop}), we finally get the correlation function
\bea
   C(\vec{x}-\vec{x}')=1 - \frac{\left({\mbox{Tr}}\left[S(\vec{x}'-\vec{x})\right]\right)^2}
   {\left({\mbox{Tr}}\left[S(0)\right]\right)^2}.
\label{corrfunc}
\eea
The denominator of the second term on the right-hand side of Eq.~(\ref{corrfunc}) is the square of the quark condensate. We consider only the thermomagnetic contribution to this condensate which is given by Eq.~(\ref{condpropthermmag}). The numerator is the square of the trace of the quark propagator in coordinate space as a function of the distance between the test particles. We thus proceed to analyze this latter object and to extract from it the change of the correlation distance when varying the temperature and field strength. 

\section{Correlation lengths}\label{IV}

In order to obtain the behavior of the correlation function when the temperature and magnetic field strength vary, we compute the trace of the quark propagator in coordinate space. For this purpose, we write the Fourier transform of Eq.~(\ref{tracewithSchwinger}) at equal times, namely,
\begin{eqnarray}
 iS(\vec{x}-\vec{x}')&=&\int\frac{d^4p}{(2\pi)^4}e^{-i(x-x')\cdot p} \nonumber \\
 &\times&\int_0^\infty \frac{ds}{\cos(q_fBs)}e^{is[p_\parallel^2-p_\perp^2\frac{\tan(q_fBs)}{q_fBs}-M^2+i\epsilon]} \nonumber \\
 &\times& \Big[ (\cos(q_fB\tau)+\gamma_1\gamma_2\sin(q_fBs))(M+\slsh{p}_\parallel)\nonumber \\
 &-&\frac{\slsh{p}_\perp }{\cos(q_fBs)} \Big]\Big|_{x_0=x_0'} .
 \label{propagatorspace}
\end{eqnarray}
We introduce finite temperature effects in Eq.~(\ref{propagatorspace}) in the manner described in Sec. II. Therefore, after performing the sum over Matsubara modes, the trace becomes
\begin{eqnarray}
 \text{Tr} [ S(\vec{x}-\vec{x}')] &=& \frac{N_c M}{4\pi^2}\frac{1}{2}\sum_f q_fB \int_{\tau_0}^\infty \frac{d\tau}{\tau \tanh(q_fB\tau)}\nonumber \\
 &\times& e^{-\tau M^2-\frac{(x_3-x'_3)^2}{4\tau}-\frac{q_fB(x_\perp-x'_\perp)^2}{4\tanh(q_fB\tau)}} \nonumber \\
 &\times& \vartheta_3\Big(\frac{1}{2},\frac{i}{4\tau \pi T^2}\Big),
 \label{traza}
\end{eqnarray}
where $x_3$ and $x_\perp$ represent the parallel and transverse (to the magnetic field) space directions, respectively, and $M\equiv M(B,T)$ is the dynamically generated mass. $\vartheta(z,x)$ is Jacobi's theta-three function that can be written as
\begin{equation}
 \vartheta_3\Big( \frac{1}{2},\frac{i}{4\pi \tau T^2} \Big)=1+2\sum_{n=1}^\infty (-1)^n e^{-\frac{n^2}{4\tau T^2}}.
 \label{jacobitheta}
\end{equation}
Substituting Eq.~(\ref{jacobitheta}) into Eq.~(\ref{traza}) we obtain
\begin{eqnarray}
 \text{Tr} [ S(\vec{x}-\vec{x}')] &=& \frac{N_c M}{4\pi^2}\frac{1}{2}\sum_f q_fB \int_{\tau_0}^\infty \frac{d\tau}{\tau \tanh(q_fB\tau)}\nonumber \\
 &\times& e^{-\tau M^2-\frac{(x_3-x'_3)^2}{4\tau}-\frac{q_fB(x_\perp-x'_\perp)^2}{4\tanh(q_fB\tau)}} \nonumber \\
 &\times& \Big \{ 1+2\sum_{n=1}^\infty (-1)^n e^{-\frac{n^2}{4\tau T^2}} \Big \}.
 \label{traza2}
\end{eqnarray}

\begin{figure}[t]
 \centering
  \includegraphics[width=0.48\textwidth]{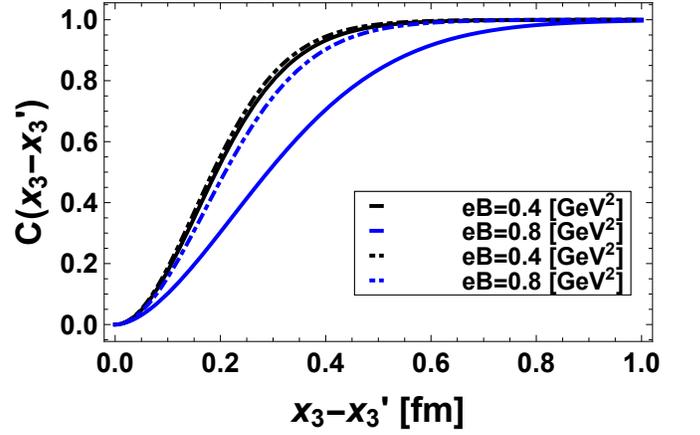}
  \caption{Correlation function in the parallel direction for $T=142$ MeV (solid lines) and at $T=178$ MeV (dashed lines) for the indicated values of the field strength $eB$. $M_0=224$ MeV.}
\label{fig2}
\end{figure}

Notice that in Eq.~(\ref{traza2}) the $T$-independent term contains both the vacuum and the pure magnetic contribution. As has been shown in Ref.~\cite{NJL2}, to isolate the vacuum contribution, we can add and subtract the $T$-independent term in the limit when $q_fB \rightarrow 0$. When doing so, we obtain
\begin{eqnarray}
 \text{Tr} [ S(\vec{x}-\vec{x}')]&=& \frac{N_c M}{4\pi^2}\frac{1}{2}\sum_f \Big \{  \int_{\tau_0}^\infty \frac{d\tau}{\tau^2} \Big( \frac{q_fB\tau}{\tanh(q_fB\tau)}
 -1 \Big)\nonumber \\
 &\times& e^{-\tau M^2-\frac{(x_3-x'_3)^2}{4\tau}-\frac{q_fB(x_\perp-x'_\perp)^2}{4\tanh(q_fB\tau)}}\nonumber \\
 &+&\int_{\tau_0}^\infty \frac{d\tau}{\tau^2}e^{-\tau M^2-\frac{(x_3-x'_3)^2}{4\tau}-\frac{q_fB(x_\perp-x'_\perp)^2}{4\tanh(q_fB\tau)}} \nonumber \\
 &+& 2q_fB \int_{\tau_0}^\infty \frac{d\tau}{\tau \tanh(q_fB\tau)}\nonumber \\
 &\times& e^{-\tau M^2-\frac{(x_3-x'_3)^2}{4\tau}-\frac{q_fB(x_\perp-x'_\perp)^2}{4\tanh(q_fB\tau)}} \nonumber \\
 &\times& \sum_{n=1}^\infty (-1)^n e^{-\frac{n^2}{4\tau T^2}} \Big \}.
 \label{trazab01}
\end{eqnarray}
Equation~(\ref{trazab01}) diverges when $x_3-x'_3=x_\perp-x'_\perp=0$. This divergence corresponds to the term we subtracted from the analysis in Eq.~(\ref{using2}). Thus, in order to isolate this term, we follow a similar procedure as described above, adding and subtracting to the integrand of Eq.~(\ref{trazab01}) the space coordinate-dependent, $T$-independent term in the limit when $q_fB \rightarrow 0$, 
\begin{eqnarray}
   \text{Tr} [ S(\vec{x}-\vec{x}')]_{B,T}&=& \frac{N_c M}{4\pi^2}\frac{1}{2}\sum_f \Big \{  \int_0^\infty \frac{d\tau}{\tau^2} 
   \nonumber \\
   &&\Big( \frac{q_fB\tau}{\tanh(q_fB\tau)}-1 \Big)\nonumber \\
  &\times& e^{-\tau M^2-\frac{(x_3-x'_3)^2}{4\tau}-\frac{q_fB(x_\perp-x'_\perp)^2}{4\tanh(q_fB\tau)}}\nonumber \\
  &+&\int_0^\infty \frac{d\tau}{\tau^2}e^{-\tau M^2} \nonumber \\
  &&\Big( e^{-\frac{(x_3-x'_3)^2}{4\tau}-\frac{q_fB(x_\perp-x'_\perp)^2}{4\tanh(q_fB\tau)}} \nonumber \\
  &-&e^{-\frac{(x_3-x'_3)^2}{4\tau}-\frac{(x_\perp-x'_\perp)^2}{4\tau}} \Big) \nonumber \\
  &+& 2q_fB \int_0^\infty \frac{d\tau}{\tau \tanh(q_fB\tau)}\nonumber \\
  &\times& e^{-\tau M^2-\frac{(x_3-x'_3)^2}{4\tau}-\frac{q_fB(x_\perp-x'_\perp)^2}{4\tanh(q_fB\tau)}} \nonumber \\
  &\times& \sum_{n=1}^\infty (-1)^n e^{-\frac{n^2}{4\tau T^2}} \Big \}.
  \label{trazab02}
 \end{eqnarray}
where we subtracted the space-dependent vacuum contribution, namely,
\bea
   \text{Tr} [ S(\vec{x}-\vec{x}')]_{0}=
   \int_{\tau_0}^\infty e^{-M^2\tau-\frac{(x_3-x'_3)^2}{4\tau}-\frac{(x_\perp-x'_\perp)^2}{4\tau}},
\label{spacevac}
\eea
and have correspondingly indicated that  Eq.~(\ref{trazab02}) contains only the thermomagnetic contribution.

\begin{figure}[t]
 \centering
  \includegraphics[width=0.48\textwidth]{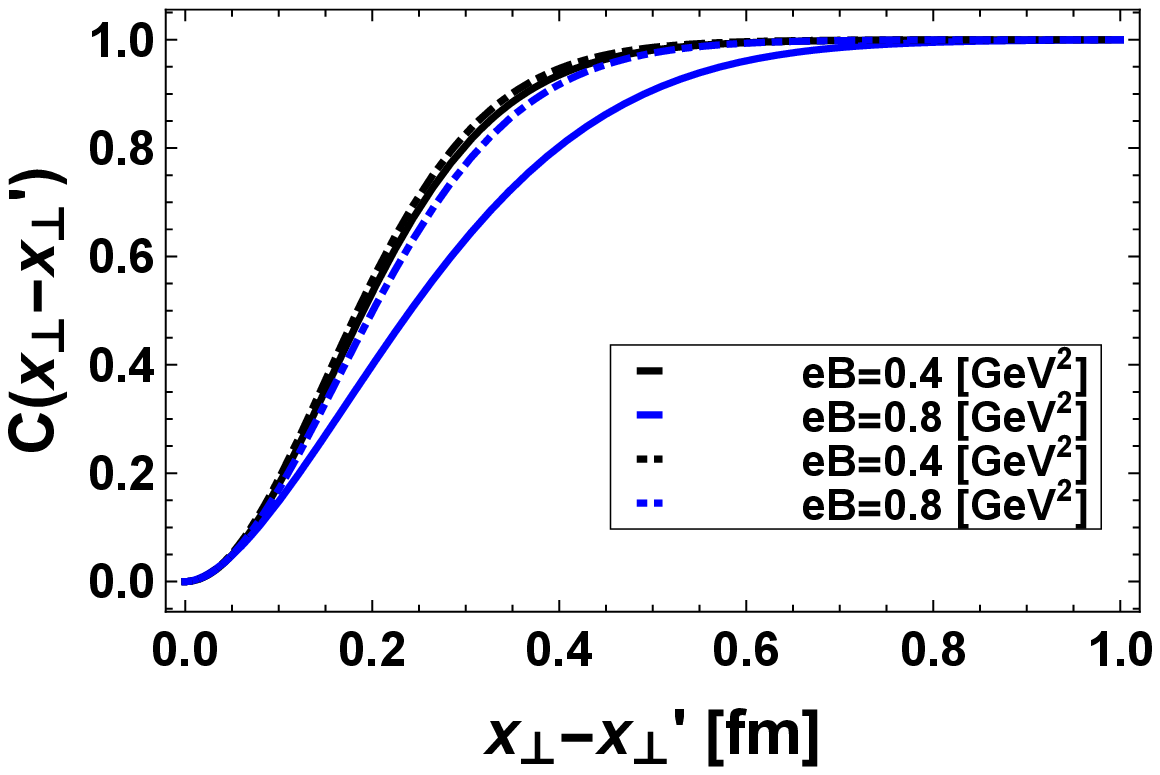}
  \caption{Correlation function in the transverse direction for $T=142$ MeV (solid lines) and at $T=178$ MeV (dashed lines) for the indicated values of the field strength $eB$. $M_0=224$ MeV.}
\label{fig3}
\end{figure}

\begin{figure}[t]
 \centering
  \includegraphics[width=0.48\textwidth]{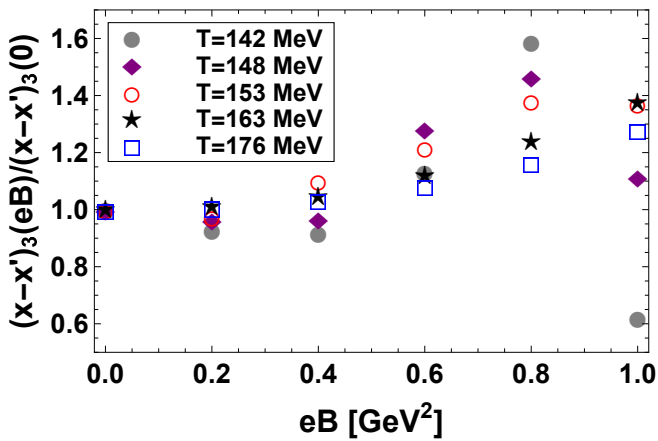}
  \caption{Correlation distance in the parallel direction as function of the field strength $eB$, normalized to the vacuum distance, with $M\equiv M(B,T)$ and for $T=142, \ 148, \ 153, \ 163 \ \text{and} \ 176$ MeV. $M_0=224$ MeV.}
\label{fig4}
\end{figure}

With Eq.~(\ref{trazab02}) at hand, we  explore the thermomagnetic behavior of the correlation function $C(\vec{x}-\vec{x}')$, Eq.~(\ref{corrfunc}).  Figures~\ref{fig2} and~\ref{fig3} show examples of the correlation function. Figure~\ref{fig2} (\ref{fig3}) shows  $C$ as a function of the longitudinal (transverse) distance between the test charges. Notice that these functions increase monotonically between 0 and 1 and that their widths depend on the values of $T$ and $B$.

In order to analyze the changes of the characteristic length contained in the correlation function, when varying $T$ and $B$, we take a fixed value of the height and study the evolution of the width of $C(\vec{x}-\vec{x}')$ for this fixed height value. We take this as $C(\vec{x}-\vec{x}')=0.5$ and define the width for this height as the correlation length. 

\begin{figure}[t]
 \centering
  \includegraphics[width=0.48\textwidth]{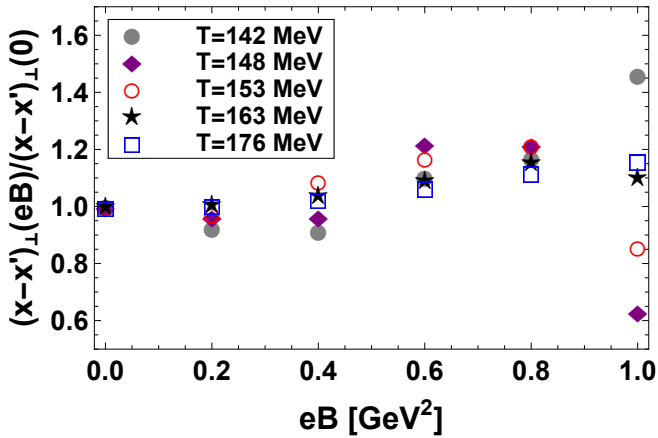}
  \caption{Correlation distance in the perpendicular direction as function of the field strength $eB$, normalized to the vacuum distance, with $M\equiv M(B,T)$ and for $T=142, \ 148, \ 153, \ 163 \ \text{and} \ 176$ MeV. $M_0=224$ MeV.}
\label{fig5}
\end{figure}

Figures~\ref{fig4} and~\ref{fig5} show the correlation lengths in the parallel and transverse directions using the values we found for $M(B,T)$ in Ref.~\cite{NJL2}, as functions of the field strength, for $T=142, \ 148, \ 153, \ 163 \ \text{and} \ 176$ MeV. For a temperature well above the pseudocritical temperature, namely, for $T=176$ MeV, both correlation distances increase monotonically with the magnetic field. As the temperature decreases from this value, there is a turn-over behavior as function of the field strength, whereby the correlation lengths start off decreasing slightly to then increase up to some value of $B$ and then decrease for larger values of $B$. The initial decrease is not as pronounced as the subsequent increase for intermediate values of $B$. 

\begin{figure}[t]
 \centering
  \includegraphics[width=0.48\textwidth]{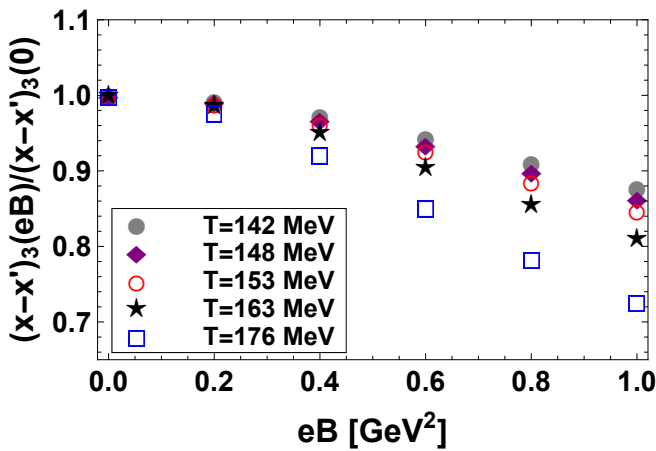}
  \caption{Correlation distance in the parallel direction as function of the field strength $eB$, normalized to the vacuum distance, with $M\equiv M(0,0)=224$ MeV and for $T=142, \ 148, \ 153, \ 163 \ \text{and} \ 176$ MeV. }
\label{fig6}
\end{figure}

\begin{figure}[t]
 \centering
  \includegraphics[width=0.48\textwidth]{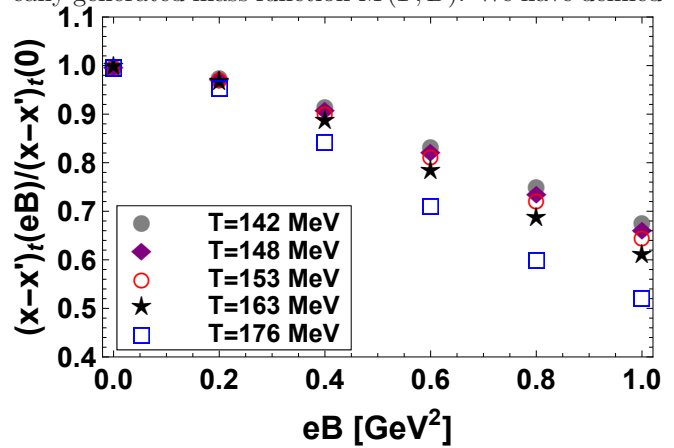}
  \caption{Correlation distance in the perpendicular direction as function of the field strength $eB$, normalized to the vacuum distance, with $M\equiv M(0,0)=224$ MeV and for $T=142, \ 148, \ 153, \ 163 \ \text{and} \ 176$ MeV.}
\label{fig7}
\end{figure}

In order to study the dependence of these correlation lengths on the properties of the dynamically generated mass $M(T,B)$, we test how these lengths are affected when considering a constant $M$. Figures~\ref{fig6} and~\ref{fig7} show this behavior as a function of $B$ for the same set of temperatures as above in the case when $M$ is constant. Notice that for all temperatures, the correlation distances decrease monotonically with the field strength.

The correlation distance exhibits very different properties when $M$ does and does not depend on $T$ and $B$. To pick the case that corresponds to the system of strongly interacting particles, recall that the behavior of the LQCD pseudocritical temperature is well reproduced by a mass function that depends on $T$ and $B$, as shown in Fig.~\ref{fig1}. This is an indication that in the NJL model, the properties of the strongly interacting magnetized plasma are also encoded in $M(T,B)$ and not only in $G(T,B)$. A constant $M$ resembles more the case of a classical system. In contrast, a full thermomagnetic dependence of $M(T,B)$ corresponds to the strongly interacting case. 

Recall that the correlation distance, as we have defined it, can be viewed as the {\it minimum distance} to place two quarks with the same quantum numbers, so that the probability for this configuration is not negligible $(C\geq 0.5)$. When this distance shrinks, for a given $T$ as a function of $B$, the system {\it tightens}. Since this happens for large $B$, when the system is below the pseudocritical temperature, we see that the magnetic interaction wins over the thermal random motion, making it to behave like the case of a classical system of charges subject to the influence of a magnetic field.  On the contrary, when the temperature is above the pseudocritical temperature, the random motion dominates such that for the explored values of $B$, the field strength is not enough to tighten up the system and, on the contrary, makes the test particles to be further apart with a large probability. It could very well be that for these temperatures a strong enough magnetic field could make the correlation distance shrink again, though this case is not considered in this work, as the field strengths are limited to values below 1 GeV$^2$ and temperatures not larger than 176 MeV. 

When the correlation distance increases, the test charges are further apart and correspondingly their interaction is weaker. To understand how the strong interaction becomes weaker as the distance between test particles increases, recall that the increase in temperature is also associated to an increase of the phase space occupation number. This means that for large temperatures more particles pop up from vacuum and these contribute to screen the interaction between test particles. When the correlation distance keeps increasing with the field strength it is as if the magnetic field brings on average one of the test particles closer to those being popped up from vacuum. Thus, on average, the increase in the occupation number together with the largest proximity between one of the test particles and those popped up from vacuum contributes to loosen up the system. This behavior is reminiscent of asymptotic freedom.

\section{Summary and conclusions}\label{concl}

In conclusion we have studied the correlation length between test quarks with the same electric and color charges in the NJL model, considering thermal and magnetic effects, encoded in the properties of the dynamically generated mass function $M(T,B)$. We have defined the correlation length from the quark correlation function which in turn is built from the probability amplitude to bring one quark into the plasma, once a previous one with the same quantum numbers is placed at a given distance apart. We found that for temperatures below the transition temperature the correlation length has a turn over behavior; it starts off growing as the field strength increases to then decrease for large magnetic fields. On the contrary, the correlation length continues increasing as the field strength increases for temperatures above the pseudocritical temperature. We have interpreted such behavior in terms of a competition between the tightening induced by the classical magnetic force and the random thermal motion. However, we have also emphasized that for large enough temperatures the increase of the occupation number contributes to the screening of the interaction between the test particles. As the correlation distance keeps growing with the magnetic field, we picture this behavior as due to the closer proximity between one of the test particles and the ones popped up from vacuum. This last behavior resembles the asymptotically free interaction between the test particle and the particles produced by the increase in temperature.   

\section*{Acknowledgements}

Support for this work has been received in part by UNAM-DGAPA-PAPIIT grant number IN101515 and by Consejo Nacional de Ciencia y Tecnolog\1a grant number 256494. R. Zamora would like to thank support from CONICYT FONDECYT Iniciaci\'on under grant No. 11160234.


\begin{thebibliography}{55}

\bibitem{bali1}G. Bali, F. Bruckmann, G. Endrodi, Z. Fodor, S. Katz,{\it et al.}, J. High Energy. Phys. {\bf 1202}, 044 (2012).

\bibitem{bali2}G. Bali, F. Bruckmann, G. Endrodi, Z. Fodor, S. Katz, {\it et al.}, Phys. Rev. D {\bf 86}, 071502 (2012).

\bibitem{bali3}G. Bali, F. Bruckmann, G. Endrodi, S. Katz, and A. Shafer, J. High Energy. Phys. {\bf 1408}, 177 (2014).

\bibitem{Boomsma01} J. K. Boomsma and D. Boer, Phys. Rev. D {\bf 81}, 074005 (2010).

\bibitem{Loewe1} M. Loewe, C. Villavicencio, ans R. Zamora, Phys. rev. D {\bf 89}, 016004 (2014).


\bibitem{Fraga} E. S. Fraga and A. Mizher, Phys. Rev. D {\bf 78}, 025016 (2008).

\bibitem{Fraga2} A. J. Mizher, M. Chernodub, and E. S. Fraga, Phys. Rev. D {\bf 82}, 105016 (2010).


\bibitem{Blaschke01} D. Blaschke, S. Frediksson, H. Grigorian, A. Oztas, and F. Sandin, Phys. Rev. D {\bf 72}, 065020 (2005).


\bibitem{Bruckmann} F. Bruckmann, G. Endr\"odi,T. G. Kovacs J. High Energy. Phys. {\bf 1304}, 112 (2013).

\bibitem{Ayala1} A. Ayala, C. A. Dom\1nguez, L. A. Hern\'andez. M. Loewe, R. Zamora, Phys. Lett. B {\bf 759}, 99-103 (2016); A. Ayala, J. J. Cobos-Mart\'inez, M. Loewe, M. E. Tejeda-Yeomans, R. Zamora, Phys. Rev. D {\bf 91}, 016007 (2015).

\bibitem{decreasing1}
R. L. S. Farias, K. P. Gomes, G. Krein and M. B. Pinto, Phys. Rev. C {\bf 90}, 025203 (2014); M. Ferreira, P. Costa, O. Louren\c{c}o, T. Frederico, C. Provid\^{e}ncia, Phys. Rev. D {\bf 89}, 116011 (2014).

\bibitem{decreasing2}
A. Ayala, M. Loewe and R. Zamora, Phys. Rev. D {\bf 91}, 016002 (2015);  A. Ayala. C. A. Dominguez, L. A. Hern\'andez, M. Loewe, R. Zamora, Phys. Rev. D {\bf 92}, 096011 (2015); A. Ayala, M. Loewe, A. J. Mizher, R. Zamora,  Phys. Rev. D {\bf 90}, 036001 (2014).

\bibitem{NJL1}
R. L. S. Farias, V. S. Timoteo, S. S. Avancini, M. B. Pinto, G. Krein, Eur. Phys. J. A {\bf 53}, 101 (2017).

\bibitem{NJL2}
A. Ayala , C. A. Dominguez, L.A. Hernandez, M. Loewe, A. Raya, J. C. Rojas, C. Villavicencio, Phys. Rev. D {\bf 94}, 054019 (2016).

\bibitem{Norberto} 
N. Scoccola, {\it private communication}.

\end{thebibliography}
\end{document}